\newcommand{\myPaperFontSize}{9pt}
\documentclass[a4paper, \myPaperFontSize, twocolumn, fleqn]{article}

\usepackage[comma,authoryear,compress]{natbib}		

\usepackage{amsmath}
\usepackage{booktabs}	
\usepackage{color}	
\usepackage{amssymb}	
\usepackage[affil-it]{authblk}					
\usepackage{csquotes}	
\usepackage[inline,shortlabels]{enumitem}	
\usepackage{graphicx}	
\usepackage{mathrsfs}	
\usepackage{stfloats}	
\usepackage{subfig}		
\usepackage{tabularx}	
\usepackage{units}		

\usepackage[hyperfootnotes=false]{hyperref}		
\usepackage[all]{hypcap}
\usepackage{xcolor}
\definecolor{dark-red}{rgb}{0.45,0.15,0.15}
\definecolor{dark-blue}{rgb}{0.15,0.15,0.4}
\definecolor{medium-blue}{rgb}{0,0,0.5}
\hypersetup{
    colorlinks, 				
    linkcolor={dark-red},		
    citecolor={dark-blue},		
    urlcolor={medium-blue},		
}

\newcommand{\protectedSubref}[1]{\protect\subref{#1}}	
\newcommand{\protectedSubrefPlain}[1]{\protect\subref*{#1}}		

\newcommand*{\fm}[1]{(\ref{#1})}	
\newcommand{\NMax}{N_\mathrm{max}}
\newcommand{\NPol}{N_\mathrm{pol}}
\newcommand{\NSS}{N_\mathrm{ss}}
\newcommand{\pEsc}{p_\mathrm{esc}}
\newcommand{\rhoLD}{\rho_\mathrm{LD}}
\newcommand{\rhoMax}{\rho_\mathrm{max}}
\newcommand{\sLD}{s_\mathrm{LD}}
\newcommand{\sMax}{s_\mathrm{max}}

\graphicspath{{.}{./figures/}}

\begin{document}


\title{Abortive Initiation as a Bottleneck for Transcription in the Early \emph{Drosophila} Embryo}


\newcommand{\affilUCLA}{Department of Chemistry And Biochemistry, University of California, Los Angeles, CA 90095, USA}
\newcommand{\affilNW}{Engineering Sciences and Applied Mathematics, Northwestern University, 2145 Sheridan Road, Evanston, IL 60208, USA}
\newcommand{\affilPasteur}{Decision and Bayesian Computation Group, Institut Pasteur, 25 Rue du Docteur Roux, Paris, 75015, France}
\author[1, 2]{Alexander S.~Serov%
\thanks{Corresponding author. E-mail: \texttt{alexander.serov@pasteur.fr}}}
\author[1]{Alexander J.~Levine}
\author[3]{Madhav Mani}
\affil[1]{\affilUCLA}
\affil[2]{\affilPasteur}
\affil[3]{\affilNW}

\maketitle

\begin{abstract}
Gene transcription is a critical step in gene expression.
The currently accepted physical model of transcription of MacDonald, Gibbs \& Pipkin~(1969) predicts the existence of a physical limit on the maximal rate of transcription, which does not depend on the transcribed gene.
This limit appears as a result of polymerase \enquote{traffic jams} forming in the bulk of the 1D DNA chain at high polymerase concentrations.
Recent experiments have, for the first time, allowed one to access live gene expression dynamics in the Drosophila fly embryo \emph{in vivo} under the conditions of heavy polymerase load and test the predictions of the model~(Garcia et al., 2013). 
Our analysis of the data shows that the maximal rate of transcription is indeed the same for the Hunchback, Snail and Knirps gap genes, and modified gene constructs in nuclear cycles~13, 14, but the experimentally observed value of the maximal transcription rate corresponds to only \unit[40]{\%} of the one predicted by this model. 
We argue that such a decrease must be due to a slower polymerase elongation rate in the vicinity of the promoter region.
This effectively shifts the bottleneck of transcription from the bulk to the promoter region of the gene.
We suggest a quantitative explanation of the difference by taking into account abortive transcription initiation.
Our calculations based on the independently measured abortive initiation constant \emph{in vitro} confirm this hypothesis and find quantitative agreement with MS2 fluorescence live imaging data in the early fruit fly embryo.
If our explanation is correct, then the transcription rate cannot be increased by replacing \enquote{slow codons} in the bulk with synonymous codons, and experimental efforts must be focused on the promoter region instead.
This study extends our understanding of transcriptional regulation, re-examines physical constraints on the kinetics of transcription and re-evaluates the validity of the standard totally asymmetric simple exclusion process model of transcription.
Several ideas on experimental validation of our hypothesis are also suggested.

\end{abstract}

\vspace{10pt}


\vspace{10pt}




\section{Introduction}

Pattern formation in the developing organism relies on precise spatiotemporal control of gene expression~\cite{Gilbert2013}.
Gene expression occurs as the result of a complex dynamical process, 
the first step of which is {\em transcription}~\cite{Alberts2002}. 
During transcription, a multi-protein polymerase complex moves sequentially down a strand of DNA~\cite{Alberts2002} and produces an mRNA molecule, which, after 
various downstream processes, produces a particular protein. 
While it is clear in some situations that the observed rate of protein production is selected by the developmental program, that rate is also subject to physical limits resulting from its molecular mechanisms~\cite{Bialek2012}.

Recently, techniques have been developed that allow us to track mRNA levels over time and across nuclear cycles of a fruit fly early development at the individual cell level~\cite{Garcia2013, Bothma2014, Bothma2015}.
Consequently, we have have access to at least parts of the fly's developmental program (as monitored via the transcription rates of a handful of specific genes) from the first cell division of the fertilized egg through
the formation of a complex and spatially patterned embryo.

These data provide a novel window on the unfolding of the fly's developmental pattern. 
It is tempting to interpret the spatiotemporal changes in the rate of mRNA transcript production solely in terms of evolutionary optimization of embryonic development.  
The development process, however, relies on physical processes, which impose their own constraints.  
In this manuscript, we examine the mRNA production data for three genes, known to be a part of the head-to-tail patterning systems active during early fly development~\cite{Gilbert2013}, with an eye towards determining whether the physical dynamics of transcription impose limitations on the developmental dynamics. 
To put this another way: we ask whether the observed developmental program runs in a dynamical regime where it saturates the physical constraints of the underlying biochemical machinery.

\begin{figure*}[htb]
\includegraphics[width = \textwidth]{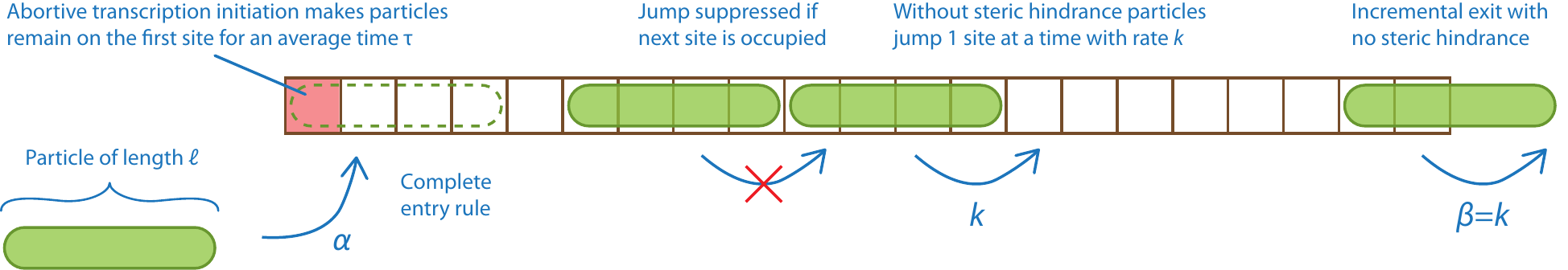}
\caption{%
Schematic representation of the abortive k-TASEP model with extended particles of length~$\ell$. 
The abortive k-TASEP model features an additional slow first site~(shown in red) compared to the standard k-TASEP model.
From the biological point of view, this model represents sequential elongation of polymerase complexes with footprint~$\ell$ along a gene of length~$L$
}
\label{fig:AbortiveTASEPScheme}
\end{figure*}

There is a distinct reason to imagine that the gene expression machinery does, in fact, saturate some dynamical upper bounds.  
On the one hand, there may exist selection pressure on the speed of embryo growth.  On the other hand, there are speed limits to transcription inherent in the mechanism itself.
In particular, the gene is a linear array of base pairs that must be read sequentially; the  polymerase complex is doing the reading by stochastically making steps along the gene, and while doing so, it takes up a finite-length footprint on it~(Fig.~\ref{fig:AbortiveTASEPScheme}).  
A number of such polymerase complexes may simultaneously run along the gene~\cite{Garcia2013}, that in a first approximation can be assumed to interact only through steric repulsion~(excluded volume). 
During maximal protein production, the number of
such complexes is large enough~(up to~100, \cite{Garcia2013}) that their combined footprint takes up a significant fraction of the gene's length~(up to~\unit[80]{\%},~\cite{Garcia2013}).
At such high coverages, one might well expect that the 
steric interactions between stochastically moving polymerase complexes set a maximal rate for gene expression. 
These basic dynamics are reminiscent of traffic jams observed on congested roadways, and has been well studied in a variety of contexts~\cite{Schutz2001, Schmittmann1995}, as we describe below.

Here we use earlier developed models to determine from theoretical considerations alone what that maximal mRNA production rate should be and ask whether this rate is indeed observed in the data.  
It is reasonable to imagine that Drosophila embryonic development is precisely the right place to look for nature to saturate this limiting value of protein production, since optimal production likely leads 
to the most efficient method for rapidly making a fly.  
Our findings are, at first, perplexing.  
We indeed find that for three different genes that are transcribed during development, the 
rates of transcriptional initiation production reach the \emph{same} maximal value~(within~\unit[15]{\%} of each other), suggesting the presence of a physically-imposed production speed limit that developmental biology saturates, at least in these extreme circumstances.  
The observed speed limit, however, is only around \unit[40]{\%} that predicted by basic theoretical 
considerations, to be described below.  
A likely explanation of this result is that we have misidentified the rate-limiting step in the mRNA production. 
One cannot discount the existence of other bottlenecks in the system, and we note here that abortive initiation of transcription has been independently observed in polymerase dynamics in the proximity of the promoter in \emph{in vitro} setups~\cite{Revyakin2006, Margeat2006, Kapanidis2006}.  
In effect, it was reported that the polymerase complex gets stuck for some time at the start of a gene. 
When we take into account this aspect of transcription and perform calculations based on an independently measured abortive initiation time constant~\cite{Revyakin2006}, we find that the traffic jams located at the start of the gene further limit the polymerase throughput and bring the predicted maximal rate inline with the observed one.

The remainder of this manuscript is organized as follows. 
First we briefly review the basic physics of traffic jams in one-dimensional models~\cite[see also][]{Schmittmann1995, Zia2011, Chou2011}. 
We then introduce a modified model based on our knowledge of abortive transcription initiation, and discuss our calculations and simulations.
The modified model is a special case of what is known in the literature as an inhomogeneous TASEP model with slow~(or defective) sites~\cite{Zia2011, Dong2009, Chou2004}.
In the next section we present the data and show that these cannot be satisfactory explained by the original model, but are quantitatively consistent with a transcriptional bottleneck associated with the initiation of the transcription process observed in the modified model. 
In the conclusions we discuss the implication of the demonstrated effect for developmental biology and for the statistical physics of non-equilibrium systems, and propose several directions for future research.

\section{The statistical mechanics of traffic jams}
Recognizing that the collective dynamics of many polymerases walking on DNA can be thought of as that of processive stochastic walkers with simple steric (excluded volume) interactions, we briefly review this long-standing model of non-equilibrium statistical physics. 
The physics of this one-dimensional model provides perhaps the simplest example of non-equilibrium steady states in a strongly interacting many-body system.  
This model, originally proposed by MacDonald and collaborators~\cite{MacDonald1968, MacDonald1969} and commonly referred to as a Totally Asymmetric Simple Exclusion Process~(TASEP), has been studied as part of a general interest in the physics of driven diffusive systems~\cite{Schmittmann1995} and 
as a model for biologically relevant dynamics~\cite{Zia2011, Chowdhury2005, Frey2004}, with the applications to ribosome motion along mRNA.
We refer the interested reader to Refs.~\cite{Schmittmann1995, Zia2011, Chou2011} for extensive reviews of the literature on TASEP.

In brief, it is essential to recall that the basic problem is defined by a particle 
injection rate~$\alpha$ at the start (e.g., the left end of the one dimensional
track on which the particles move) and an extraction rate at the right end~(Fig.~\ref{fig:AbortiveTASEPScheme}). 
The injection rate~$\alpha$ may be interpreted as the attempt frequency to add a particle to the left, which is successful only when that space is unoccupied. 
The extraction rate~$\beta$ may be similarly interpreted. 
Once injected, particles move stochastically to the right.  
We will consider only the case of Poisson walkers taking single steps on a lattice so that, without interparticle interactions, a single stepping rate~$k$ determines both the mean velocity of the walker 
and all its velocity fluctuations~\cite{vanKampen1992}.  
We also restrict our analysis to the case in which particles cannot be added or removed in the bulk~(i.e. no Langmuir kinetics), although novel dynamics have been studied in this case~\cite{Parmeggiani2003}.  
This restriction is demanded by the experimentally observed processivity of the polymerase complex on the DNA.  
The model contains two more control parameters: the length of the track~$L$~(i.e., the total number of base pairs in the gene) and the length of the walking particles~$\ell$~(i.e., the number of lattice sites taken up by one walker).
In most TASEP studies the walkers are assumed to be \enquote{point particles}~(taking up one site) on a lattice constant in length, but, as we will see below, the finite length of the polymerase complex plays a significant role in understanding the collective dynamics of transcription in the developing embryo and cannot be ignored.

The main phenomenology of~TASEP dynamics of point-like particles can be characterized by a phase diagram defined by the injection and extraction rates $\alpha$
and $\beta$ respectively~\cite{Schutz2001, Zia2011}.  
It is remarkable that this one-dimensional system with short-ranged interactions can develop long-range order such that the bulk properties of the system can be controlled by these boundary conditions at the endpoints. 
This is one of the many features distinguishing non-equilibrium steady states from the better understood thermodynamic phases of equilibrium states. 
A mean-field analysis of the steady states leads to the prediction of three distinct phases~\cite{Schutz2001}.  
For~$\alpha, \beta$ sufficiently small, one finds a stable \emph{shock front} where the particle density jumps from a bulk value defined by the bulk jump rate~$k$ to the one controlled by the boundary extraction/injection rates~$\alpha$,~$\beta$.  
The position of this stable shock within the system depends on the 
relative values of~$\alpha$ and~$\beta$.  
For sufficiently high values of both $\alpha$ and $\beta$, one observes a different phase, the \emph{maximal current regime}. For~$\alpha > \beta$~($\beta > \alpha$) 
there is also a high (low) density phase in which the current through the system is below the maximal value. 
These mean-field results are supported by exact solutions to the steady-state
density profile obtained by \citeauthor{Liggett1975}~\cite{Liggett1975} and by numerical simulations.

In the case of extended particles~(the k-TASEP model), the conclusions of the mean-field approach stay valid, but additional factors appear in now approximate formulas for the mean elongation rate~$s$ and particle density~$\rho$.
For instance, in the low-density~(LD) phase for low injection attempt rate~$\alpha$ and high exit attempt rate~$\beta$ one obtains~\cite{MacDonald1969, Shaw2003}:
\begin{align}
\label{fm:LowDensityPhaseOriginal}
\rhoLD(\alpha) = \frac{\alpha}{k+\alpha(\ell-1)}, %
&&\sLD(\alpha) = \frac{\alpha(k - \alpha)}{k + \alpha(\ell - 1)}.
\end{align}
When the injection attempt rate~$\alpha$ increases to its critical value %
$\chi = k/(1+\sqrt{l}),$ %
while~$\beta$ stays high~($\beta\geqslant \chi$), the system experiences a phase transition to the maximal current phase that is characterized by the maximal particle density~$\rhoMax$ and the maximal particle elongation rate~$\sMax$:
\begin{align}
\label{fm:MaximalCurrentPhaseOriginal}
\rhoMax = \frac{1}{\sqrt{\ell}(1 + \sqrt{\ell})}, %
&&\sMax = \frac{k}{(1 + \sqrt{l})^2}.
\end{align}
For more details on the phase diagram of the k-TASEP model, the reader is referred to Refs~\cite{Shaw2003, Schutz2001}.
Assuming the footprint of a polymerase complex on the DNA is $\ell=45\pm5$ base pairs~(bp) long~(as determined by electron crystallography,~\citep{Rice1993, Selby1997, Poglitsch1999}), and its bulk jump rate on the~DNA is around $k\approx\unit[26\pm2]{bp/s}$~\citep{Garcia2013}, in the maximal current regime, the theory predicts the maximal polymerase injection rate~$\sMax\approx\unit[0.44\pm0.05]{pol/s}\approx\unit[26\pm3]{pol/min}$.
New polymerase complexes cannot be recruited on the given gene faster than this rate, which is defined by how fast the injected particle clears the injection site~(the first~$\ell$ sites of the lattice).
On a gene~$L=\unit[5400]{bp}$ base pairs long~(like \emph{Hunchback} gene, \citep{Garcia2013}), the maximal number of polymerases is~$\NMax \equiv L\rhoMax\approx \unit[104\pm11]{pol}$.

\section{Abortive k-TASEP Model}

In this paper we suggest that abortive initiation of transcription can be the rate-limiting step of transcription and should be added to the classical k-TASEP model, when used to explain experimental data in early embryo development. 
One can expect that in such model the bottleneck of the transcription process will be located in the promoter region at the 5' end of the gene, rather than in its bulk as in the original model. 
Abortive initiation is an effect wherein polymerase complexes loaded on a promoter region of a gene will several times engage in transcription cycles, but will abort and release short~RNA products returning to the initial position~\citep{Kapanidis2006, Revyakin2006, Margeat2006}.
However once a product of~9 to~11 nucleotides is synthesized, the polymerase will break its interactions with the promoter and will proceed with mRNA synthesis.
It is currently accepted that the transition from the abortive initiation state to processive~RNA synthesis occurs through a \enquote{DNA scrunching} mechanism, when the mechanical stress stored in the \enquote{scrunched} DNA during abortive transcription allows the polymerase to break interactions with the promoter.
However other mechanisms~(namely, \enquote{transient excursions}, \enquote{inchworming}) have been proposed as well~\citep{Kapanidis2006}.

To assess the effect of abortive initiation on the transcription rate, we modify the standard k-TASEP model by adding a slow site at the beginning of the lattice. 
This modification will cause the polymerase to stochastically pause at the first site imitating the effect observed in the experiment.
Experimental studies have revealed that the time spent by a polymerase in the abortive initiation stage in the promoter region is an exponentially-distributed random variable with the average pause time~$\tau\approx\unit[5\pm1]{s}$~\citep{Revyakin2006}, which we will use in the simulations below.
In what follows, we will call this modification of the k-TASEP model an \emph{abortive k-TASEP model}.

\begin{figure*}[tb]
\newcommand{\myScale}{0.32}
\centering \subfloat[]{
\includegraphics[width = \myScale\textwidth]
{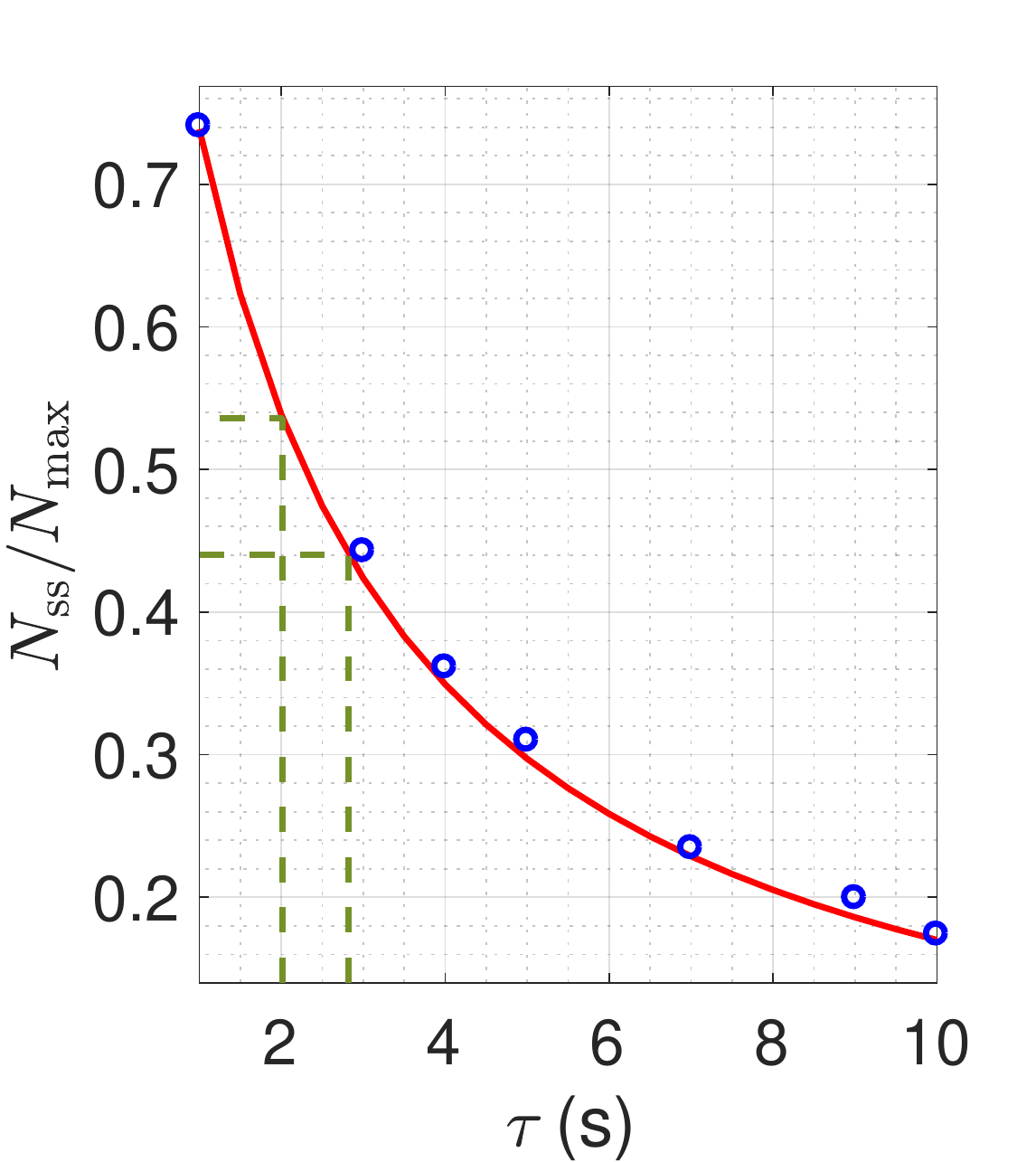}
\label{fig:SimulationMaxNumber}  
}
\centering \subfloat[]{
\includegraphics[width = \myScale\textwidth]
{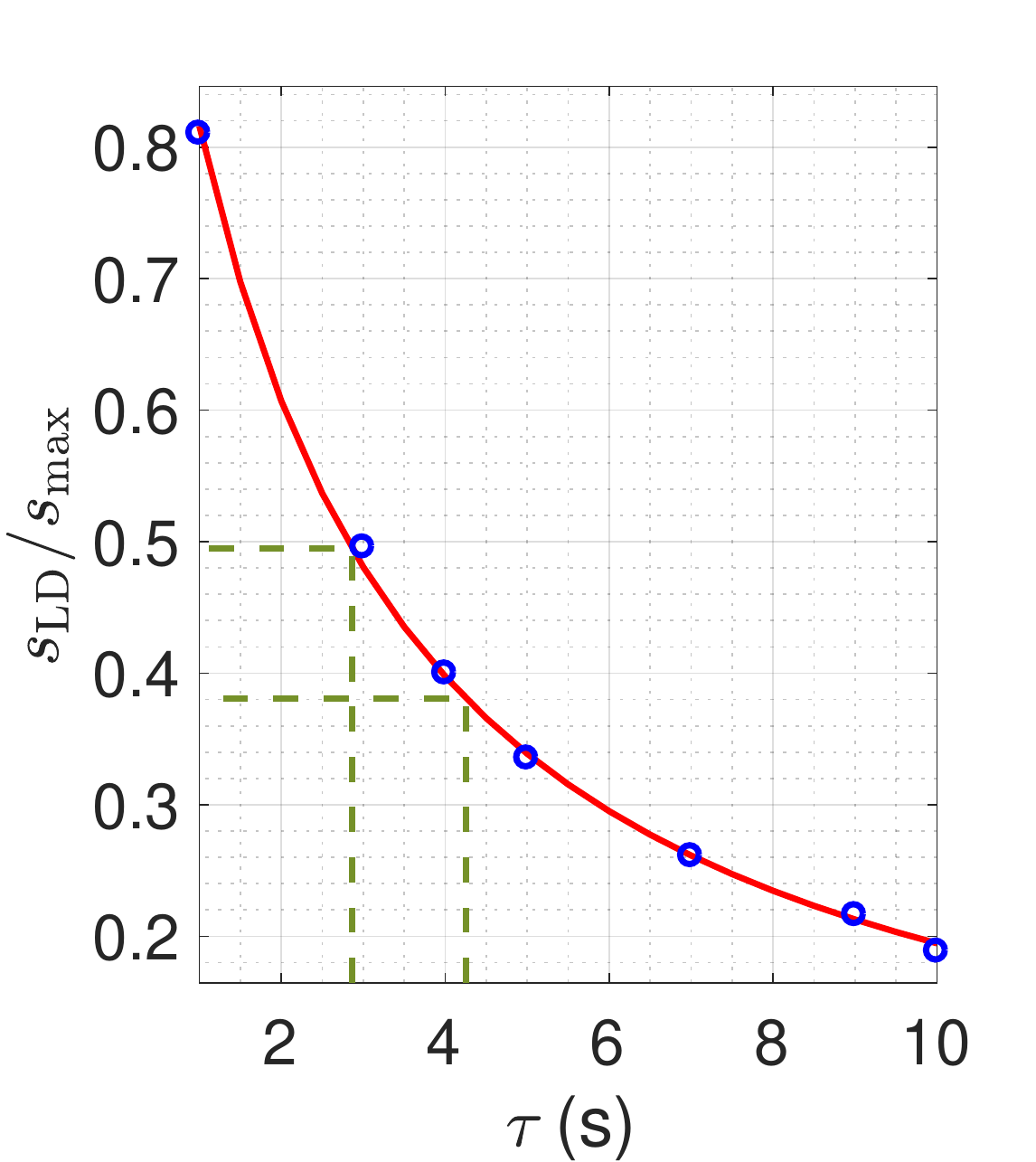}
\label{fig:SimulationInjectionRate}  
}
\centering \subfloat[]{
\includegraphics[width = \myScale\textwidth]{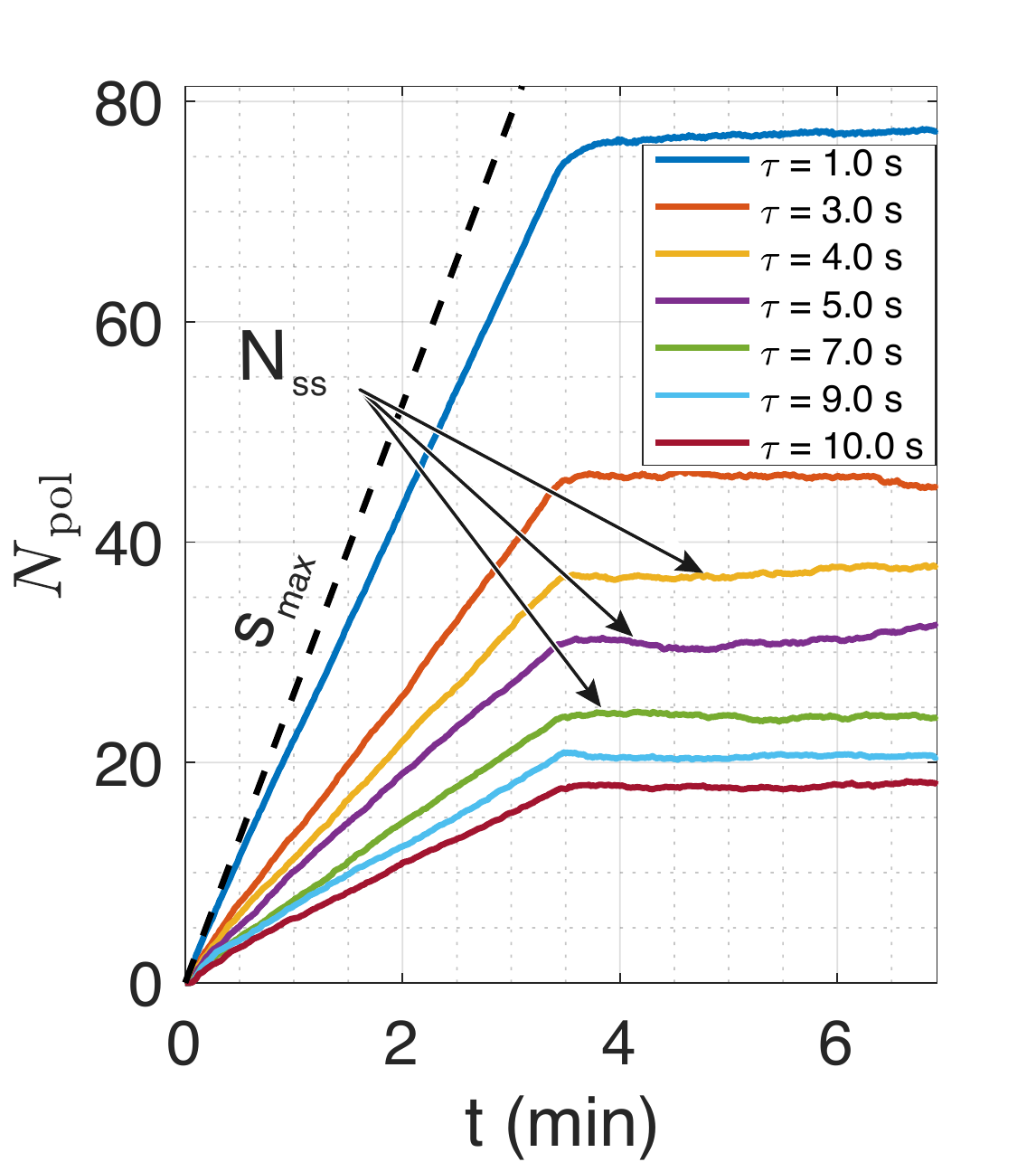}
\label{fig:SimulationMeanEvolution}
} 
\centerline{
}
\caption{%
Numerical simulations of %
the maximal polymerase number in the steady state~$\NSS$~\protectedSubref{fig:SimulationMaxNumber}, %
the polymerase injection rate~$s$~\protectedSubref{fig:SimulationInjectionRate}, %
and the evolution of polymerase number~$\NPol$~\protectedSubref{fig:SimulationMeanEvolution} %
as functions of the mean time~$\tau$ spent in the abortive initiation regime.
Each circle in~\protectedSubref{fig:SimulationMaxNumber},\protectedSubref{fig:SimulationInjectionRate} and each curve in~\protectedSubref{fig:SimulationMeanEvolution} represent the average of~20 simulations; the dashed black line marks the maximal loading rate~$\sMax\approx\unit[26\pm3]{pol/min}$ predicted by the original k-TASEP model~(Eq.~\fm{fm:MaximalCurrentPhaseOriginal}).
Independent experiments report that the mean duration of the abortive initiation phase is around~$\tau\approx\unit[5\pm1]{s}$~\citep{Revyakin2006}, which gives $\NSS/\NMax\approx0.30$ and $s/\sMax\approx0.34$, or $\NSS\approx31$ and $s\approx\unit[8.8]{pol/min}$ in the absolute values.
Black arrows in~\protectedSubref{fig:SimulationMeanEvolution} indicate the established steady state of the simulation where~$\NSS$ is evaluated.
$s$~corresponds to the slope of the curves in~\protectedSubref{fig:SimulationMeanEvolution}, with~$\sMax$ labeled in the plot.
Eq.~\fm{fm:LowDensityPhaseWithTau} and plots~\protectedSubref{fig:SimulationMaxNumber} and \protectedSubref{fig:SimulationInjectionRate} can be used to assess the mean time~$\tau$ polymerases spend in the abortive initiation regime based on experimental data for each gene, nc and construct.
For instance, for the Hunchback bac construct~(Fig.~\ref{fig:Histograms}), one gets $\tau\approx\unit[2.0]{s}$ in nc~13 and $\tau\approx\unit[2.8]{s}$ in nc~14 from the mean $\NSS$, and $\tau\approx\unit[2.9]{s}$ in nc~13 and $\tau\approx\unit[4.3]{s}$ in nc~14 from the mean injection rate~$s$~(dashed green lines).
}
\label{fig:Simulations}
\end{figure*}

The proposed abortive k-TASEP model is a special case of an inhomogeneous k-TASEP model with only one slow site located at the first site of the lattice.
The inhomogeneous TASEP model has attracted significant interest in recent years, and some important analytical and numerical results have been obtained for point particles~
\cite[see][and references therein]{Cook2013, Poker2015, Dhiman2016, Xiao2016}, %
as well as for extended particles~\cite{Shaw2003, Shaw2004, Shaw2004a, Dong2007, Klumpp2008, Dong2009, Zia2011, Brackley2011}.
The standard way to treat a problem with an individual slow site like ours would consist in using mean-field approximations for the sublattices to the left and to the right of the slow site and stitching the two solutions together by the flux conservation equation on the slow site~\cite{Kolomeisky1998, Shaw2004a}.
However, when the only slow site is located at the beginning of the lattice, the problem can be easily reduced to the standard k-TASEP model~(a similar problem with point particles was recently considered by~\cite{Xiao2016}).
Indeed, the slow site can be considered as a boundary condition for the remaining~$L-1$ sites of the lattice that can be treated as an ordinary~$(L-1)$-site k-TASEP model.
Ignoring entry-rule-related boundary effects, one can then, in the first approximation, define the injection attempt rate~$\alpha$ of the rest of the lattice through the pause time~$\tau$ at the first site:~$\alpha = 1/\tau$.
For long genes~$L\gg1$, one can safely assume~$L-1\approx L$ and using Eq.~\fm{fm:LowDensityPhaseOriginal}, \fm{fm:MaximalCurrentPhaseOriginal} get the following expressions for the normalized polymerase number in the steady state and normalized polymerase elongation rate:
\begin{equation}
\label{fm:LowDensityPhaseWithTau}
\begin{aligned}
\NSS / \NMax &= \frac{\sqrt{\ell}(1+\sqrt{\ell})}{k\tau + \ell - 1}, %
\\
\sLD / \sMax &= \frac{(k\tau -1)(1 + \sqrt{\ell})^2}{k\tau (k\tau + \ell - 1)}.
\end{aligned}
\end{equation}
These dependencies are shown in Figs.~\ref{fig:SimulationMaxNumber},\protectedSubrefPlain{fig:SimulationInjectionRate} as solid red lines for the biologically relevant range of values of~$\tau=\unit[1\text{--}10]{s}$.
Note that the normalized polymerase number~$\NSS$ and the normalized injection rate~$\sLD$ decrease from~$\NSS / \NMax\approx0.74$ and~$s/\sMax\approx0.82$ for~$\tau=\unit[1]{s}$ to~$\NSS / \NMax\approx0.17$ and~$s/\sMax\approx0.19$ for~$\tau=\unit[10]{s}$.
For the experimentally reported value of~$\tau\approx\unit[5\pm1]{s}$~\cite{Revyakin2006}, the corresponding ratios are~$\NSS / \NMax\approx0.30$ and~$s/\sMax\approx0.34$.

Note that while $\NSS/\NMax$ describes the steady-state of the system, the expression for~$\sLD/\sMax$ describes the current related to a shock wave of density propagating in the system before the steady state is established.
The shock-wave description~\cite{Schutz2001} is possible thanks to the fact that the interface separating two distinct phases in the TASEP model is microscopic, i.e. the transition occurs on a scale of several lattice sites~\cite{Janowsky1992, Janowsky1994}.
We verify that the shock-wave approach and the approximate formulas~\fm{fm:LowDensityPhaseWithTau} ignoring boundary-layer effects correctly describe our abortive k-TASEP model by performing Monte Carlo simulations with the numerical values of parameters~$\ell$, $L$ and~$k$ provided above.
We use the \enquote{complete entry, incremental exit} rule implying that polymerases are loaded on the gene only if the first~$\ell$ sites are free, while they leave the gene one site at a time with the bulk rate~$k$~(Fig.~\ref{fig:AbortiveTASEPScheme})~\citep{Lakatos2003}.
To focus on the effect of the first slow site, we set the polymerase injection rate~$p$ to the value~$p=\unit[100]{s^{-1}}$ much greater than the inverse time a polymerase spends on the first site of the lattice~$p\gg 1/\tau$ for all~$\tau$ in the range~\unit[1--10]{s}. 
The slow site is modeled as a stochastic process, in which at each time step~$\Delta t$ a polymerase can either remain at the first~$\ell$ sites or continue transcription with a certain finite probability.
The escape probability~$\pEsc=\Delta t/\tau$~(the probability to leave the slow site) is calculated based on the chosen values of~$\Delta t$ and~$\tau$.
The simulations were performed with the \emph{rejection-free kinetic Monte Carlo} algorithm~(KMC, an event-based algorithm, also known as \emph{residence-time} or \emph{Bortz-Kalos-Lebowitz} algorithm, see~\citep{Bortz1975}).
The theoretical prediction~\fm{fm:LowDensityPhaseWithTau} is found to perfectly describe simulation results~(shown as blue circles in Figs~\ref{fig:SimulationMaxNumber},\protectedSubrefPlain{fig:SimulationInjectionRate}), which validates our analytical approach.
Fig.~\ref{fig:SimulationMeanEvolution} completes the simulation results illustrating the mean evolution of the system from the non-steady state to the steady state.


\section{Comparison to Experimental Data}
In the present section we compare the predictions of the abortive k-TASEP model to experimental results inferred from MS2 fluorescence data after~FISH calibration, as described in~\citep{Garcia2013}.
Each data entry~(fluorescence trace) contains the number of RNA polymerase~II complexes~$N(t)$ currently loaded~(transcribing) on a selected gene as a function of time~$t$ recorded in one nucleus of one Drosophila fly embryo.
The whole observation period is divided into nuclear cycles~(nc), at the end of each of which the number of cells in the embryo doubles, and all polymerase complexes are unloaded from the gene.
We define the time point~$t=0$ to correspond to the start of nc~10~\citep{Garcia2013}.
In this analysis we will consider only the nuclear cycles~13~(starting around~$t\approx\unit[35]{min}$) and~14~(starting around~$t\approx\unit[50]{min}$), since the genes we analyze are mainly expressed in these nuclear cycles.

The following three genes were considered in these study: Hunchback~(Hb), Snail~(Sn) and Knirps~(Kn).
All considered genes have one primary and one shadow enhancer that influence the spatio-temporal dynamics of expression of these genes.
The data presented below were obtained for each of these genes in three modifications: wild-type~(labeled \emph{bac}), a gene with no primary enhancer~(\emph{no primary}) and a gene with no shadow enhancer~(\emph{no shadow}).
Table~\ref{tbl:FluorescenceDataSummary} summarizes the number of embryos and fluorescence traces analyzed for each gene-enhancer construct combination.
The time resolution of the measurements is~\unit[37]{s}.

\begin{table}[tbp]
\newcommand{\tblWidth}{\columnwidth}
\caption{
Number of fluorescence traces analyzed for each gene-enhancer construct combination.
The number of embryos is given in the parentheses
}
	\label{tbl:FluorescenceDataSummary}
		\begin{tabularx}{\tblWidth}{@{}l*{3}{>{\centering\arraybackslash}X}@{}}
		\toprule
		&	Hunchback	&	Snail	&	Knirps
		\\
		\midrule
		Bac	&	1965 (11)	&	547 (5)	&	326 (5)
		\\
		No primary	&	979 (9)	&	607 (4)	&	84 (2)		
		\\
		No shadow	&	1279 (8)	&	159 (1)	&	60 (1)
		\\
		\bottomrule
		\end{tabularx}
\end{table}

\begin{figure}[tb]
\newcommand{\myScale}{0.5}
\centering \subfloat[]{
\includegraphics[width = 0.5\columnwidth]
{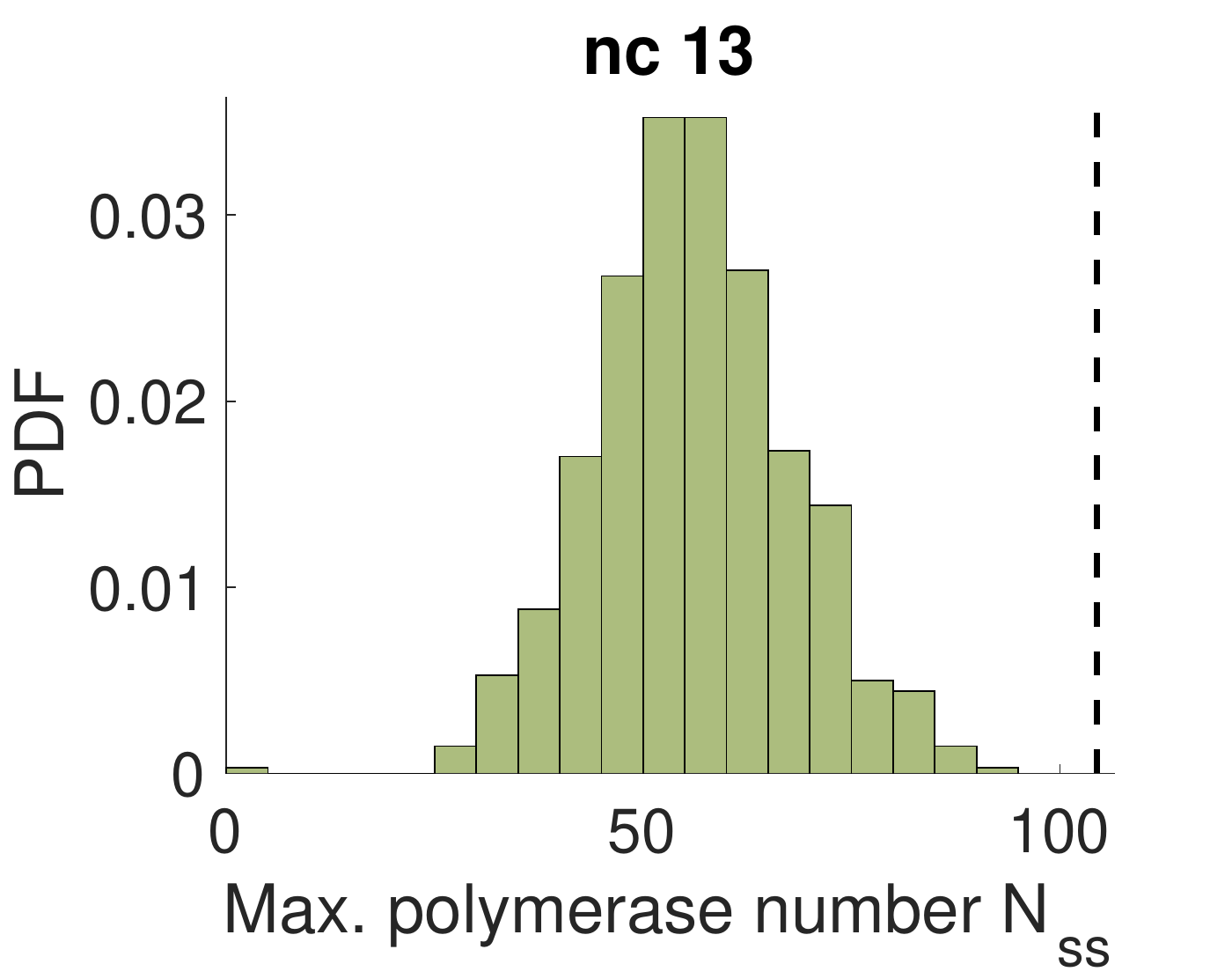}
\label{fig:MaxNumberHistogramsNC13}  
}
\centering \subfloat[]{
\includegraphics[width = 0.5\columnwidth]
{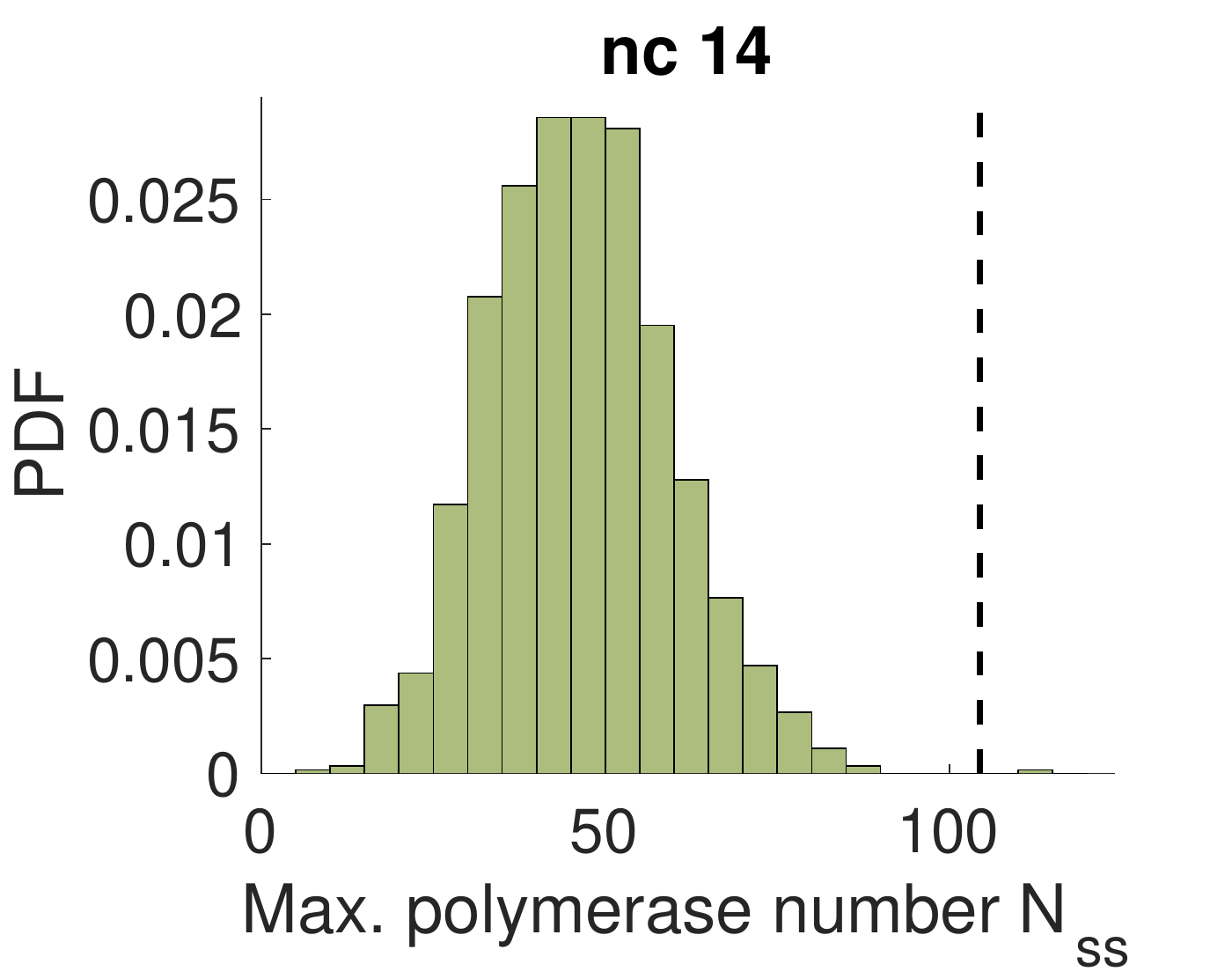}
\label{fig:MaxNumberHistogramsNC14}  
}
\\
\centering \subfloat[]{
\includegraphics[width = 0.5\columnwidth]
{01b_1_HunchBack_max_numbers_histograms_bac.pdf}
\label{fig:SlopeHistogramsNC13}  
}
\centering \subfloat[]{
\includegraphics[width = 0.5\columnwidth]
{01b_2_HunchBack_max_numbers_histograms_bac.pdf}
\label{fig:SlopeHistogramsNC14}  
}
\caption{
\protectedSubref{fig:MaxNumberHistogramsNC13},\protectedSubref{fig:MaxNumberHistogramsNC14}:~Probability density function~(PDF) of the steady-state number of polymerases~$\NSS$ loaded on the HunchBack bac gene in nuclear cycles~13~\protectedSubref{fig:MaxNumberHistogramsNC13} and~14~\protectedSubref{fig:MaxNumberHistogramsNC14} (all AP positions combined).
All data entries are grouped in bins of width~5, and the y-axis is proportional to the number of fluorescence traces falling in the current bin.
The theoretically predicted upper limit~$\NMax\approx \unit[104\pm11]{pol}$ is shown by a dashed black line.
The mean values of the distributions in the figure are: $\NSS=\unit[56\pm12]{pol}$ (nc 13) and $\NSS=\unit[46\pm13]{pol}$ (nc 14).
\protectedSubref{fig:SlopeHistogramsNC13},\protectedSubref{fig:SlopeHistogramsNC14}:~PDF of initial polymerase injection rates~$s$ for the HunchBack bac enhancer construct in nuclear cycles 13~\protectedSubref{fig:SlopeHistogramsNC13} and 14~\protectedSubref{fig:SlopeHistogramsNC14} (all AP positions combined).
All slopes are grouped in bins of width~1, and the y-axis is proportional to the number of fluorescence traces falling in the current bin.
The dashed vertical line shows the maximal polymerase injection rate~$\sMax\approx\unit[26]{pol/min}$ as predicted by Eq.~\fm{fm:MaximalCurrentPhaseOriginal}.
The mean values of the shown distributions are: $s=\unit[13]{pol/min}$ (nc 13) and $s=\unit[10]{pol/min}$ (nc 14).
In all figures the total number of traces shown is~682 for nc~13, and~1283 for nc~14.
One can see that both~$\NMax$ and~$\sMax$ correctly predict the maximal observed values, although the average values are lower.
The few outliers above~$\sMax$ and~$\NSS$ are likely due to noisy experimental data
%
}
\label{fig:Histograms}
\end{figure}

\begin{figure}[tb]
\centerline{
\includegraphics[width = \columnwidth]{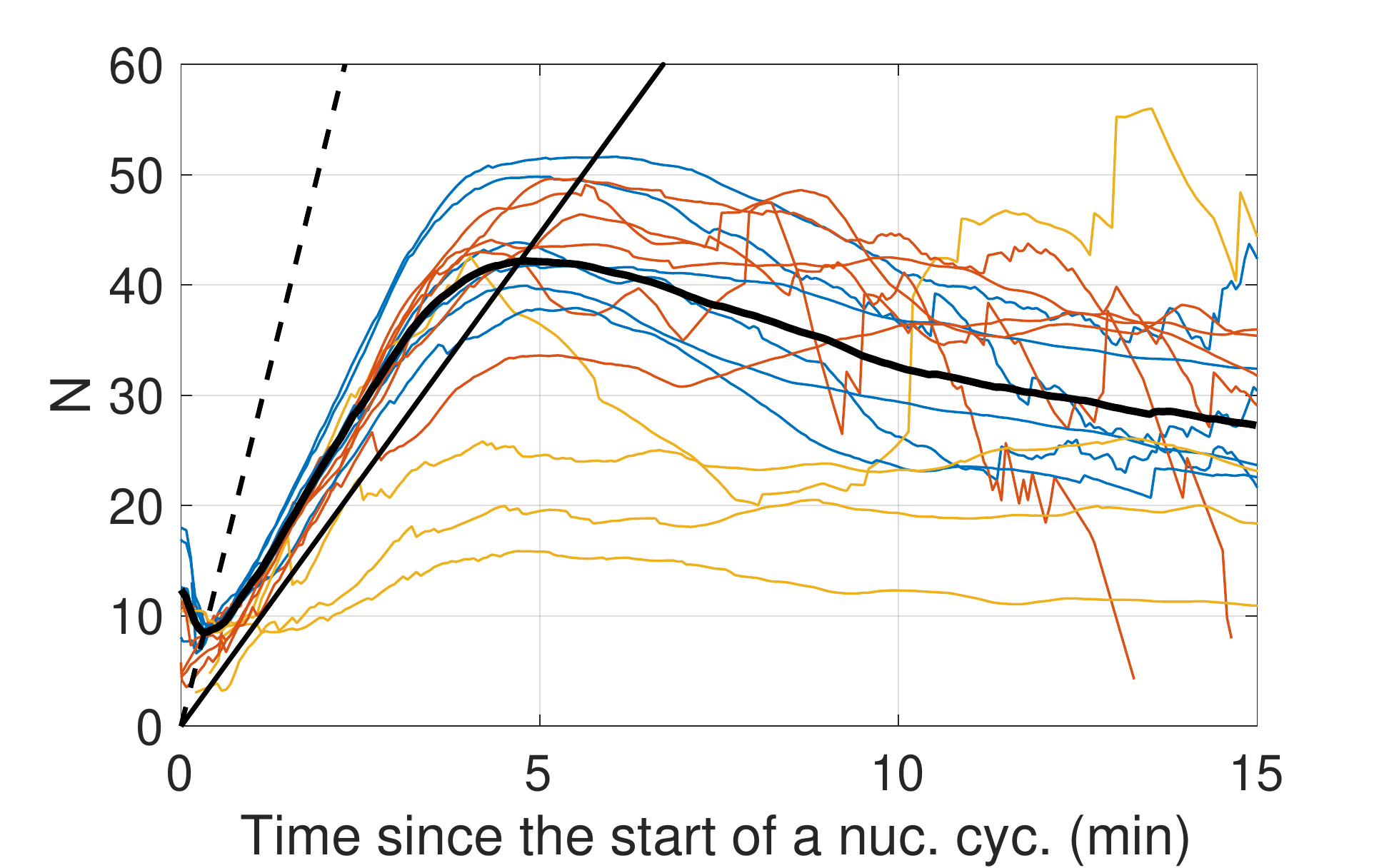}
}
\caption{
Mean evolution of active polymerase number on the Hunchback~(blue), Snail~(yellow) and Knirps~(red) genes.
Each curve corresponds to a specific construct~(bac, no primary or no shadow) of a particular gene~(Hunchback, Snail or Knirps) in one of the nuclear cycles~(13 or 14) and is averaged over the anterior-posterior~(AP) position in the embryo.
The curves are combined in the same plot to demonstrate that the initial injection rate~$s$ does not depend on any of these parameters~(some noise is present).
The thick black curve shows the average over all the parameters.
The dashed straight line shows the maximal theoretical injection rate~$\sMax\approx\unit[26]{pol/min}$ as predicted by the standard k-TASEP model~(Eq.~\fm{fm:MaximalCurrentPhaseOriginal}).
The solid straight line shows the initial injection rate~$s/\sMax\approx 0.34$ as predicted by the abortive k-TASEP model for~$\tau\approx\unit[5]{s}$.
Eqs~\fm{fm:LowDensityPhaseWithTau} can be used to adjust the~$\tau$ parameter and fit the initial slope more accurately~(see main text).
All fluorescence traces used throughout this article required synchronization of the beginning of nuclear cycles between different embryos and nuclei.
The desynchronization is due to recording started at different moments in different embryos and to the limited temporal resolution of the data.
All traces shorter than \unit[10]{mins} were also removed, since we could not reliably synchronize them with the other traces.
The steady state of the system seems to be evolving with time, showing a gradual decrease in the active polymerase number.
Under these conditions, we have taken the peak number of polymerases as an estimate of~$\NSS$ in all presented data
}
\label{fig:MeanEvolution}
\end{figure}

\begin{figure}[tb]
\centerline{
\includegraphics[width = \columnwidth]
{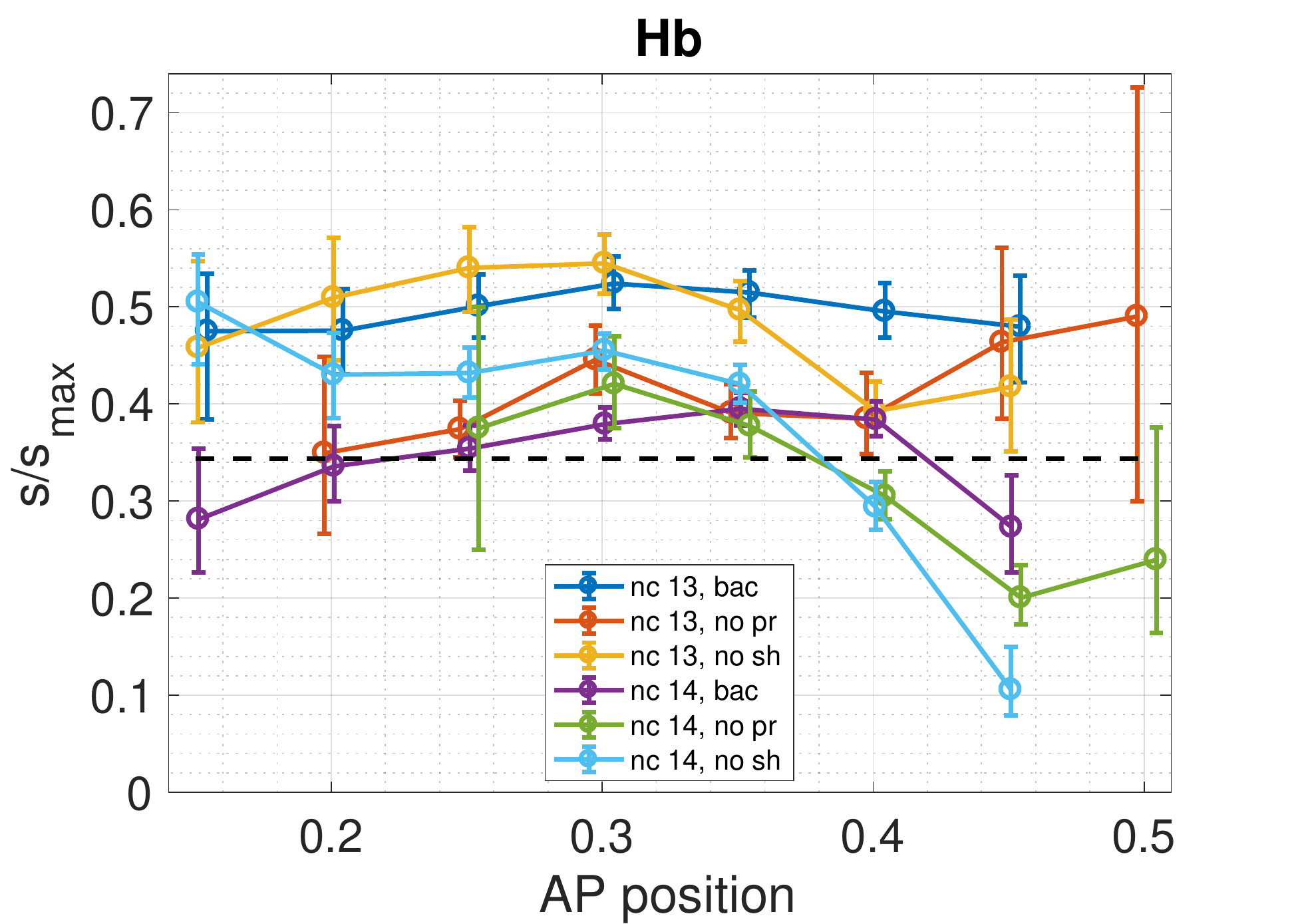}
}
\caption{Initial injection rate~$s$ as a function of the AP~position for different constructs~(bac, no primary, no shadow) of the Hunchback gene. 
The error bars show \unit[95]{\%} confidence intervals as estimated by bootstrapping~\citep{Efron1979}.
Each curve was slightly shifted sideways to allow all error bars to be visible.
The dashed line shows the injection rate predicted by the abortive TASEP model~($s/\sMax\approx0.34$) for~$\tau=\unit[5]{s}$.
This prediction fits well the experimental data lying slightly below.
Narrow AP windows and low number of data sets available to us for the Snail and Knirps genes~(Table~\ref{tbl:FluorescenceDataSummary}) did not allow us to draw any reliable conclusions on the dependence of~$s/\sMax$ on~AP for those genes
}
\label{fig:SlopeVsAP}
\end{figure}

Figs~\ref{fig:SlopeHistogramsNC13},\protectedSubrefPlain{fig:SlopeHistogramsNC14} show the distribution of polymerase loading rates~$s$ as measured in our experiments for \emph{bac Hb} in nc~13 and~14. 
The maximal theoretical polymerase injection rate~$\sMax\approx\unit[26\pm3]{pol/min}$ predicted by the original k-TASEP model is marked by dashed black lines.
One can see that~$\sMax$ appears to envelope all observed experimental values, suggesting that Hb transcription in nc~13 and~14 cannot occur faster than~$\sMax$.
The steady-state polymerase number~$\NSS$ for the Hunchback bac gene is shown in Figs~\ref{fig:MaxNumberHistogramsNC13},\protectedSubrefPlain{fig:MaxNumberHistogramsNC14}, with the k-TASEP prediction~$\NMax \approx \unit[104\pm11]{pol}$ shown by a dotted black line.
Once again we see that~$\NMax$ provides a correct overall limit for all observed experimental values, suggesting that the original k-TASEP model indeed defines the upper boundary on the number of polymerase complexes loaded on the gene in nc~13 and~14.
Moreover, Figs~\ref{fig:MeanEvolution} and~\ref{fig:SlopeVsAP} demonstrate that the actual mean transcription rate~$s$ does not depend on a particular gene, construct or location of the nucleus on the anterior-to-posterior~(AP) axis of the embryo, suggesting gene-independent nature of the transcription rate constraint.

It is however clear that the mean values of the distributions in~Fig.~\ref{fig:Histograms} lie well below the predictions of the unmodified k-TASEP model~\cite{MacDonald1969, Shaw2003}.
The histograms in Fig.~\ref{fig:Histograms} clearly demonstrate a downshift from the predicted numbers, indicating that on average the 1D DNA chain is used significantly lower than its maximum polymerase flow capacity for a homogeneous lattice~(with, for instance,~$\NSS/\NMax \approx 0.48$ and~$s/\sMax\approx0.44$ for \emph{Hb bac}, Fig.~\ref{fig:Histograms}).
In this paper we interpret this decrease as that on average no traffic jams are observed in the bulk of the~DNA chain for the selected genes and constructs, and that for them abortive transcription initiation is the real bottleneck of transcription~(other interpretations are also possible, and we return to this discussion later).
Our speculations are supported by the abortive k-TASEP model that for independently measured value of~$\tau\approx\unit[5\pm1]{s}$~\citep{Revyakin2006} predicts a similar decrease:~$\NSS / \NMax\approx0.30$ and~$s/\sMax\approx0.34$.
The slight difference between the predicted and the observed values of~$\NSS/\NMax$ and~$s/\sMax$ may be~(besides other reasons such as biological variability) attributed to a slightly different value of~$\tau$ for Drosophila embryos \emph{in vivo} compared to previous \emph{in vitro} experiments in Ref.~\cite{Revyakin2006}.
If one assumes the validity of the abortive k-TASEP model, the distributions in Fig.~\ref{fig:Histograms} can be explained by the following values of~$\tau$: $\tau\approx\unit[2.0]{s}$ for nc~13 and $\tau\approx\unit[2.8]{s}$ for nc~14 from the polymerase number~$\NSS$, and $\tau\approx\unit[2.9]{s}$ for nc~13 and $\tau\approx\unit[4.3]{s}$ for nc~14 from the mean injection rate~$s$.
These values are shown by dashed green lines in Figs~\ref{fig:SimulationMaxNumber},\protectedSubrefPlain{fig:SimulationInjectionRate}.

\begin{figure*}[tb]
\centerline{
\includegraphics[width = \textwidth]{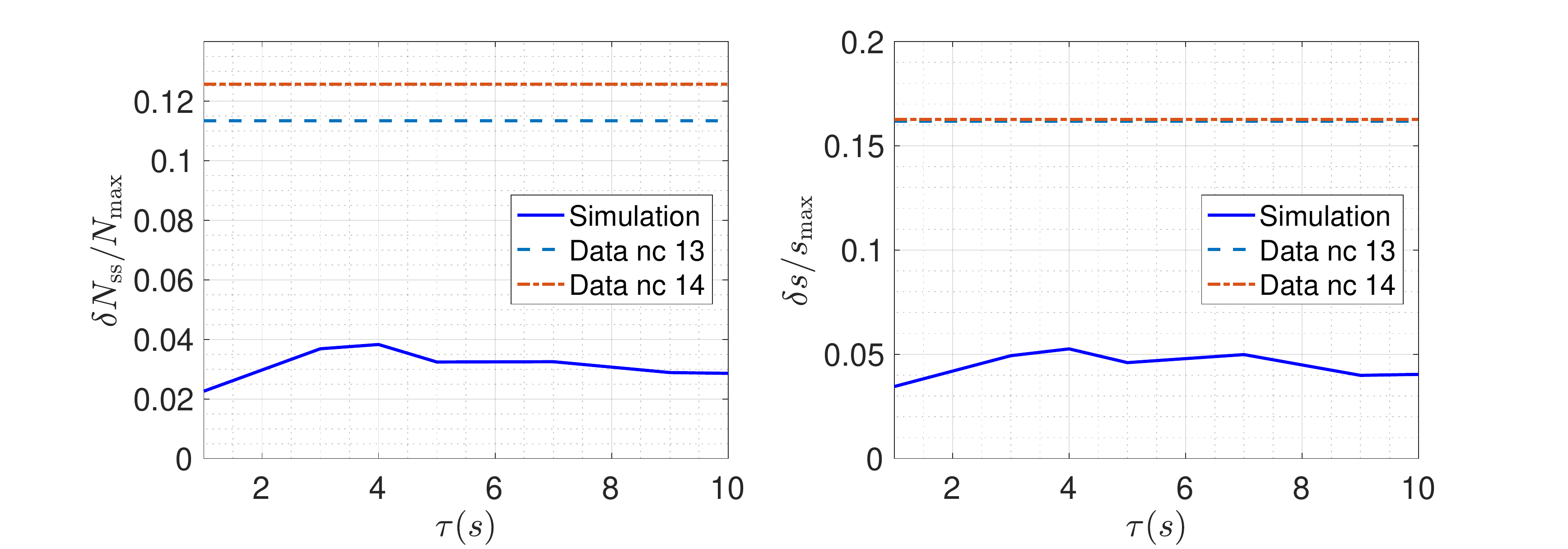}
}
\caption{Relative standard deviation of the steady state number of polymerases~($\delta \NSS/\NMax$) and normalized initial injection slope~($\delta s/\sMax$) calculated numerically as a function of mean abortive transcription initiation duration~$\tau$.
Experimentally calculated values for the Hunchback bac construct are shown by dashed horizontal lines: $\delta \NSS/\NMax\approx 0.115$~(nc~13), $\delta \NSS/\NMax\approx 125$~(nc~14), $\delta s/\sMax\approx 0.16$~(nc~13), $\delta s/\sMax\approx 0.16$~(nc~14).
The experimental lines show a constant value, since~$\tau$ dependence was not available from the experiment.
Note that the noise level of the experimental data is much higher than the noise predicted by the numerical model, which suggests the existence of additional noise sources in experimental data~(see main text)
}
\label{fig:STDComparison}
\end{figure*}

To finalize the comparison between the simulations and the experimental data, it is interesting to compare the widths of the calculated distributions for~$\NSS$ and~$s$ to the widths of experimental distributions in Fig.~\ref{fig:Histograms}.
Fig.~\ref{fig:STDComparison} shows standard deviations~(STD) of $\NSS/\NMax$ and $s/\sMax$ calculated for different values of mean abortive initiation time~$\tau$ and plotted along the experimentally measured values. 
It can be seen that experimental distributions are about~3 times wider than the calculated ones, which suggests that the noise sources included into the numerical abortive k-TASEP model~(namely, stochastic elongation and stochastic abortive initiation duration) are responsible for only about one third of the variability of the experimental values.
Other noise sources~(such as stochastic promoter activation, heterogeneous or dynamic elongation rates, instrumental noise or slope fitting) must account for the rest of the width of the distributions~(for further discussion of noise sources in protein synthesis see~\citep{Tkacik2008}).

\section{Discussion}

Our analysis of transcriptional dynamics in the Drosophila fly embryo exhibits a common value of steady-state polymerase number~$\NSS$ and polymerase injection rate~$s$ in the beginning of nuclear cycles~13 and~14 for several analyzed genes and constructs.
Although such universal character can be explained by the standard k-TASEP model, the observed values of~$\NSS$ and~$s$ are significantly lower than the theoretically predicted ones. 
In the present article we have demonstrated that this difference in slopes can be explained by the abortive initiation process that effectively decreases the polymerase elongation rate in the promoter region of the gene.
We have modified the k-TASEP model adding a slow first site, and provided simple formulas and performed Monte Carlo simulations to estimate~$\NSS$ and~$s$ in it.
Our results for independently measured abortive transcription initiation duration~$\tau\approx\unit[5]{s}$ give values much closer to the experimental ones, and suggest that the promoter region may be the real bottleneck of the transcription process rather than polymerase \enquote{traffic jams} downstream of the gene.

It is however important to mention that our hypothesis is not the only possible explanation for the observed phenomenon.
It seems to be impossible to explain the observed decrease in~$s/\sMax$ and~$\NSS/\NMax$ without slow sites near the start of the gene, but the distribution of slow sites may take different forms.
We have only analyzed the simplest option of only one first slow site.
Other options may include several distributed slow sites, clusters of slow sites~\cite{Chou2004}, or inhomogeneous jump rates~(quenched disorder) near the start of the gene~\cite{Shaw2004, Zia2011, Shaw2003}.
Interestingly, the effect of quenched disorder on the particle current in the TASEP model has been shown to mainly reduce to the effect of several slowest sites~\cite{Shaw2004, Shaw2003}.
Although the physical description of systems with different distributions of slow sites would be similar, the biological processes behind the slow sites may be completely different.
For instance, the quenched disorder model describes the fact that different nucleotides for the building mRNA chain may have different concentrations in the cell, so that the polymerase complex has to wait longer for some nucleotides than for the others thus modulating the elongation speed.
It is interesting to mention here that, for example, for some of the proteins of \emph{Escherichia coli}, the slow codons were reported to be preferentially located within the first 25~codons~\cite{Chen1990}.
Regardless of the exact distribution of slow sites, an important practical conclusion of this study is that if the bottleneck of transcription is indeed located in the promoter region, the overall transcription rate cannot be increased by modifications in the bulk of the gene, such as replacement of slow codons with synonymous ones~\cite{Shaw2004, Zia2011, Shaw2003}.
Experimental research of higher transcription rates must then focus on acceleration of the initiation dynamic or on substitution of slow codons in the promoter region of the gene.

From the physical point of view, it seems to be impossible to distinguish different configurations of slow sites having only access to the steady-state dynamics of the experimental system.
However, one can probably search for the answer in the non-steady state dynamics, or the slow evolution of the steady-states seen in Fig.~\ref{fig:MeanEvolution}.
For instance, if slow sites are distributed along the whole gene, the propagating density wave must slow down each time a new, slower site is discovered.
Surprisingly, the curves in Fig.~\ref{fig:MeanEvolution} don't seem to exhibit this gradual or stepwise slow-down, suggesting again that the slowest sites are located near the promoter region.
This feature, as well as a local decrease of transcription speed around~\unit[1--2]{mins} into nc~13~(Fig.~\ref{fig:MeanEvolution}, middle pannel), or the gradual decrease of polymerase density after the maximum has been reached~(Fig.~\ref{fig:MeanEvolution}), cannot be explained by the abortive k-TASEP model, and require further investigation.

From the point of view of molecular biology, one could conceive additional experiments aimed at identification of the role of the abortive transcription process in the decrease of the transcription rate.
On the one hand, if one could alter the molecular mechanisms behind the abortive initiation and slow it down, the corresponding decrease in the transcription rate and steady-state polymerase number could be observed experimentally and compared to the predictions of Eq.~\fm{fm:LowDensityPhaseWithTau} for the new values of~$\tau$.
If on the other hand, the abortive initiation could be accelerated, at some point one can expect that the transcription bottleneck may shift to other involved mechanisms, so that the transcription rate may be determined by polymerase \enquote{traffic jams} in the bulk of the gene or, for instance, promoter activation kinetics.
In either case, the curves in Fig.~\ref{fig:MeanEvolution} must reflect the changes.

We have also compared the width of the distributions of~$\NSS/\NMax$ and~$s/\sMax$ in our simulations and in the experimental data.
Much wider experimental distributions indicate that the noise sources taken into account in our abortive k-TASEP model cannot satisfactory explain the full variability of experimental data, and that additional noise sources should be investigated in future studies.
As an interesting development, one could try to extract additional information about the transcriptional bottlenecks and noise sources of the system by analyzing the noise, identifying noise signatures seen in the experimental data and shapes of the experimental distributions~(i.e., in Fig.~\ref{fig:Histograms}) and comparing them to those predicted by the abortive k-TASEP model or other noise sources.

Finally, the question of whether the system of polymerases on the DNA actually reaches a steady non-equilibrium state under physiological conditions of nc~13 and~14 also requires additional investigation.
For instance, Fig.~\ref{fig:MeanEvolution} is inconclusive about whether the transcription rate stays constant for some time after reaching the peak value and before starting to decrease.
If it is not the case, the validity of the steady-state k-TASEP model as a model of transcription under physiological conditions should be critically re-evaluated.

\bibliography{..//library.bib}

\begin{thebibliography}{42}
\providecommand{\natexlab}[1]{#1}
\providecommand{\url}[1]{\texttt{#1}}
\expandafter\ifx\csname urlstyle\endcsname\relax
  \providecommand{\doi}[1]{doi: #1}\else
  \providecommand{\doi}{doi: \begingroup \urlstyle{rm}\Url}\fi

\bibitem[Alberts et~al.(2002)Alberts, Johnson, Lewis, Raff, Roberts, and
  Walter]{Alberts2002}
Alberts,~B., Johnson,~A., Lewis,~J., Raff,~M., Roberts,~K., and Walter,~P.,
  2002.
\newblock \emph{{Molecular Biology of the Cell}}.
\newblock Garland Science, New York, 4 edition.

\bibitem[Bialek(2012)]{Bialek2012}
Bialek,~W., 2012.
\newblock \emph{{Biophysics: Searching for Principles}}.
\newblock Princeton University Press, Princeton, Oxford.

\bibitem[Bortz et~al.(1975)Bortz, Kalos, and Lebowitz]{Bortz1975}
Bortz,~A., Kalos,~M., and Lebowitz,~J.~L., 1975.
\newblock {A new algorithm for Monte Carlo simulation of Ising spin systems}.
\newblock \emph{J. Comput. Phys.}, 17\penalty0 (1):\penalty0 10--18.
\newblock \url{http://dx.doi.org/10.1016/0021-9991(75)90060-1}.

\bibitem[Bothma et~al.(2014)Bothma, Garcia, Esposito, Schlissel, Gregor, and
  Levine]{Bothma2014}
Bothma,~J.~P., Garcia,~H.~G., Esposito,~E., Schlissel,~G., Gregor,~T., and
  Levine,~M., 2014.
\newblock {Dynamic regulation of eve stripe 2 expression reveals
  transcriptional bursts in living Drosophila embryos}.
\newblock \emph{Proc. Natl. Acad. Sci. U. S. A.}, 111\penalty0 (29):\penalty0
  10598--603.
\newblock \url{http://dx.doi.org/10.1073/pnas.1410022111}.

\bibitem[Bothma et~al.(2015)Bothma, Garcia, Ng, Perry, Gregor, and
  Levine]{Bothma2015}
Bothma,~J.~P., Garcia,~H.~G., Ng,~S., Perry,~M.~W., Gregor,~T., and Levine,~M.,
  2015.
\newblock {Enhancer additivity and non-additivity are determined by enhancer
  strength in the Drosophila embryo}.
\newblock \emph{Elife}, 4\penalty0 (AUGUST2015):\penalty0 1--14.
\newblock \url{http://dx.doi.org/10.7554/eLife.07956.001}.

\bibitem[Brackley et~al.(2011)Brackley, Romano, and Thiel]{Brackley2011}
Brackley,~C.~A., Romano,~M.~C., and Thiel,~M., 2011.
\newblock {The Dynamics of Supply and Demand in mRNA Translation}.
\newblock \emph{PLoS Comput. Biol.}, 7\penalty0 (10):\penalty0 e1002203.
\newblock \url{http://dx.doi.org/10.1371/journal.pcbi.1002203}.

\bibitem[Chen and Inouye(1990)]{Chen1990}
Chen,~G. F.~T. and Inouye,~M., 1990.
\newblock {Suppression of the negative effect of minor arginine codons on gene
  expression; preferential usage of minor codons within the first 25 codons of
  the Escherichia coli genes}.
\newblock \emph{Nucleic Acids Res.}, 18\penalty0 (6):\penalty0 1465--1473.
\newblock \url{http://dx.doi.org/10.1093/nar/18.6.1465}.

\bibitem[Chou and Lakatos(2004)]{Chou2004}
Chou,~T. and Lakatos,~G., 2004.
\newblock {Clustered bottlenecks in mRNA translation and protein synthesis}.
\newblock \emph{Phys. Rev. Lett.}, 93\penalty0 (19):\penalty0 1--4.
\newblock \url{http://dx.doi.org/10.1103/PhysRevLett.93.198101}.

\bibitem[Chou et~al.(2011)Chou, Mallick, and Zia]{Chou2011}
Chou,~T., Mallick,~K., and Zia,~R. K.~P., 2011.
\newblock {Non-equilibrium statistical mechanics: from a paradigmatic model to
  biological transport}.
\newblock \emph{Reports Prog. Phys.}, 74\penalty0 (11):\penalty0 116601.
\newblock \url{http://dx.doi.org/10.1088/0034-4885/74/11/116601}.

\bibitem[Chowdhury et~al.(2005)Chowdhury, Schadschneider, and
  Nishinari]{Chowdhury2005}
Chowdhury,~D., Schadschneider,~A., and Nishinari,~K., 2005.
\newblock {Physics of transport and traffic phenomena in biology: From
  molecular motors and cells to organisms}.
\newblock \emph{Phys. Life Rev.}, 2\penalty0 (4):\penalty0 318--352.
\newblock \url{http://dx.doi.org/10.1016/j.plrev.2005.09.001}.

\bibitem[Cook et~al.(2013)Cook, Dong, and Lafleur]{Cook2013}
Cook,~L.~J., Dong,~J.~J., and Lafleur,~A., 2013.
\newblock {Interplay between finite resources and a local defect in an
  asymmetric simple exclusion process}.
\newblock \emph{Phys. Rev. E - Stat. Nonlinear, Soft Matter Phys.}, 88\penalty0
  (4):\penalty0 1--8.
\newblock \url{http://dx.doi.org/10.1103/PhysRevE.88.042127}.

\bibitem[Dhiman and Gupta(2016)]{Dhiman2016}
Dhiman,~I. and Gupta,~A.~K., 2016.
\newblock {Origin and dynamics of a bottleneck-induced shock in a two-channel
  exclusion process}.
\newblock \emph{Phys. Lett. Sect. A Gen. At. Solid State Phys.}, 380\penalty0
  (24):\penalty0 2038--2044.
\newblock \url{http://dx.doi.org/10.1016/j.physleta.2016.04.031}.

\bibitem[Dong et~al.(2007)Dong, Schmittmann, and Zia]{Dong2007}
Dong,~J.~J., Schmittmann,~B., and Zia,~R. K.~P., 2007.
\newblock {Inhomogeneous exclusion processes with extended objects: The effect
  of defect locations}.
\newblock \emph{Phys. Rev. E - Stat. Nonlinear, Soft Matter Phys.}, 76\penalty0
  (5):\penalty0 31--33.
\newblock \url{http://dx.doi.org/10.1103/PhysRevE.76.051113}.

\bibitem[Dong et~al.(2009)Dong, Zia, and Schmittmann]{Dong2009}
Dong,~J.~J., Zia,~R. K.~P., and Schmittmann,~B., 2009.
\newblock {Understanding the edge effect in TASEP with mean-field theoretic
  approaches}.
\newblock \emph{J. Phys. A Math. Theor.}, 42\penalty0 (1):\penalty0 015002.
\newblock \url{http://dx.doi.org/10.1088/1751-8113/42/1/015002}.

\bibitem[Efron(1979)]{Efron1979}
Efron,~B., 1979.
\newblock {Bootstrap Methods: Another Look at the Jackknife}.
\newblock \emph{Ann. Stat.}, 7\penalty0 (1):\penalty0 1--26.
\newblock \url{http://dx.doi.org/10.1214/aos/1176344552}.

\bibitem[Frey et~al.(2004)Frey, Parmeggiani, and Franosch]{Frey2004}
Frey,~E., Parmeggiani,~A., and Franosch,~T., 2004.
\newblock {Collective phenomena in intracellular processes}.
\newblock \emph{Genome informatics}, 15\penalty0 (1):\penalty0 46--55.
\newblock \url{http://dx.doi.org/10.11234/gi1990.15.46}.

\bibitem[Garcia et~al.(2013)Garcia, Tikhonov, Lin, and Gregor]{Garcia2013}
Garcia,~H.~G., Tikhonov,~M., Lin,~A., and Gregor,~T., 2013.
\newblock {Quantitative imaging of transcription in living Drosophila embryos
  links polymerase activity to patterning.}
\newblock \emph{Curr. Biol.}, 23\penalty0 (21):\penalty0 2140--5.
\newblock \url{http://dx.doi.org/10.1016/j.cub.2013.08.054}.

\bibitem[Gilbert(2013)]{Gilbert2013}
Gilbert,~S.~F., 2013.
\newblock \emph{{Developmental Biology}}.
\newblock Sinauer Associates, Inc., Sunderland, MA, 10 edition.

\bibitem[Janowsky and Lebowitz(1994)]{Janowsky1994}
Janowsky,~S.~A. and Lebowitz,~J.~L., 1994.
\newblock {Exact results for the asymmetric simple exclusion process with a
  blockage}.
\newblock \emph{J. Stat. Phys.}, 77\penalty0 (1-2):\penalty0 35--51.
\newblock \url{http://dx.doi.org/10.1007/BF02186831}.

\bibitem[Janowsky and Lebowitz(1992)]{Janowsky1992}
Janowsky,~S.~A. and Lebowitz,~J.~L., 1992.
\newblock {Finite-size effects and shock fluctuations in the asymmetric
  simple-exclusion process}.
\newblock \emph{Phys. Rev. A}, 45\penalty0 (2):\penalty0 618--625.
\newblock \url{http://dx.doi.org/10.1103/PhysRevA.45.618}.

\bibitem[Kapanidis et~al.(2006)Kapanidis, Margeat, Ho, Kortkhonjia, Weiss, and
  Ebright]{Kapanidis2006}
Kapanidis,~A.~N., Margeat,~E., Ho,~S.~O., Kortkhonjia,~E., Weiss,~S., and
  Ebright,~R.~H., 2006.
\newblock {Initial transcription by RNA polymerase proceeds through a
  DNA-scrunching mechanism.}
\newblock \emph{Science (80-. ).}, 314\penalty0 (5802):\penalty0 1144--1147.
\newblock \url{http://dx.doi.org/10.1126/science.1131399}.

\bibitem[Klumpp and Hwa(2008)]{Klumpp2008}
Klumpp,~S. and Hwa,~T., 2008.
\newblock {Stochasticity and traffic jams in the transcription of ribosomal
  RNA: Intriguing role of termination and antitermination.}
\newblock \emph{Proc. Natl. Acad. Sci. U. S. A.}, 105\penalty0 (47):\penalty0
  18159--18164.
\newblock \url{http://dx.doi.org/10.1073/pnas.0806084105}.

\bibitem[Kolomeisky(1998)]{Kolomeisky1998}
Kolomeisky,~A.~B., 1998.
\newblock {Asymmetric simple exclusion model with local inhomogeneity}.
\newblock \emph{J. Phys. A. Math. Gen.}, 31\penalty0 (4):\penalty0 1153--1164.
\newblock \url{http://dx.doi.org/10.1088/0305-4470/31/4/006}.

\bibitem[Lakatos and Chou(2003)]{Lakatos2003}
Lakatos,~G. and Chou,~T., 2003.
\newblock {Totally asymmetric exclusion processes with particles of arbitrary
  size}.
\newblock \emph{J. Phys. A. Math. Gen.}, 36\penalty0 (8):\penalty0 2027--2041.
\newblock \url{http://dx.doi.org/10.1088/0305-4470/36/8/302}.

\bibitem[Liggett(1975)]{Liggett1975}
Liggett,~T.~M., 1975.
\newblock {Ergodic theorems for the asymmetric simple exclusion process}.
\newblock \emph{Trans. Am. Math. Soc.}, 213:\penalty0 237--61.
\newblock \url{http://dx.doi.org/10.1090/S0002-9947-1975-0410986-7}.

\bibitem[MacDonald and Gibbs(1969)]{MacDonald1969}
MacDonald,~C.~T. and Gibbs,~J.~H., 1969.
\newblock {Concerning the kinetics of polypeptide synthesis on polyribosomes}.
\newblock \emph{Biopolymers}, 7\penalty0 (5):\penalty0 707--725.
\newblock \url{http://dx.doi.org/10.1002/bip.1969.360070508}.

\bibitem[MacDonald et~al.(1968)MacDonald, Gibbs, and Pipkin]{MacDonald1968}
MacDonald,~C.~T., Gibbs,~J.~H., and Pipkin,~A.~C., 1968.
\newblock {Kinetics of biopolymerization on nucleic acid templates}.
\newblock \emph{Biopolymers}, 6\penalty0 (1):\penalty0 1--25.
\newblock \url{http://dx.doi.org/10.1002/bip.1968.360060102}.

\bibitem[Margeat et~al.(2006)Margeat, Kapanidis, Tinnefeld, Wang, Mukhopadhyay,
  Ebright, and Weiss]{Margeat2006}
Margeat,~E., Kapanidis,~A.~N., Tinnefeld,~P., Wang,~Y., Mukhopadhyay,~J.,
  Ebright,~R.~H., and Weiss,~S., 2006.
\newblock {Direct observation of abortive initiation and promoter escape within
  single immobilized transcription complexes.}
\newblock \emph{Biophys. J.}, 90\penalty0 (4):\penalty0 1419--1431.
\newblock \url{http://dx.doi.org/10.1529/biophysj.105.069252}.

\bibitem[Parmeggiani et~al.(2003)Parmeggiani, Franosch, and
  Frey]{Parmeggiani2003}
Parmeggiani,~A., Franosch,~T., and Frey,~E., 2003.
\newblock {Phase coexistence in driven one-dimensional transport.}
\newblock \emph{Phys. Rev. Lett.}, 90\penalty0 (February):\penalty0 086601.
\newblock \url{http://dx.doi.org/10.1103/PhysRevLett.90.086601}.

\bibitem[Poglitsch et~al.(1999)Poglitsch, Meredith, Gnatt, Jensen, Chang, Fu,
  and Kornberg]{Poglitsch1999}
Poglitsch,~C.~L., Meredith,~G.~D., Gnatt,~a.~L., Jensen,~G.~J., Chang,~W.~H.,
  Fu,~J., and Kornberg,~R.~D., 1999.
\newblock {Electron crystal structure of an RNA polymerase II transcription
  elongation complex.}
\newblock \emph{Cell}, 98\penalty0 (6):\penalty0 791--798.
\newblock \url{http://dx.doi.org/S0092-8674(00)81513-5 [pii]}.

\bibitem[Poker et~al.(2015)Poker, Margaliot, and Tuller]{Poker2015}
Poker,~G., Margaliot,~M., and Tuller,~T., 2015.
\newblock {Sensitivity of mRNA Translation.}
\newblock \emph{Sci. Rep.}, 5:\penalty0 12795.
\newblock \url{http://dx.doi.org/10.1038/srep12795}.

\bibitem[Revyakin et~al.(2006)Revyakin, Liu, Ebright, and Strick]{Revyakin2006}
Revyakin,~A., Liu,~C., Ebright,~R.~H., and Strick,~T.~R., 2006.
\newblock {Abortive Initiation and Productive Initiation by RNA Polymerase
  Involve DNA Scrunching}.
\newblock \emph{Science (80-. ).}, 314\penalty0 (5802):\penalty0 1139--43.
\newblock \url{http://dx.doi.org/10.1126/science.1131398}.

\bibitem[Schmittmann and Zia(1995)]{Schmittmann1995}
Schmittmann,~B. and Zia,~R.
\newblock {Statistical Mechanics of Driven Diffusive Systems}.
\newblock In Domb,~C. and Lebowitz,~J.~L., editors, \emph{Phase Transitions
  Crit. Phenomena. Vol. 17}, pages 3--214. Academic Press, London, 1995.
\newblock \url{http://dx.doi.org/10.1016/S1062-7901(06)80014-5}.

\bibitem[Sch{\"{u}}tz(2001)]{Schutz2001}
Sch{\"{u}}tz,~G.~M.
\newblock {Exactly Solvable Models for Many-Body Systems Far from Equilibrium}.
\newblock In Domb,~C. and Lebowitz,~J.~L., editors, \emph{Phase Transitions
  Crit. Phenom.}, chapter~1, pages 3--255. Academic Press, San Diego, 2001.
\newblock \url{http://dx.doi.org/10.1016/S1062-7901(01)80015-X}.

\bibitem[Selby et~al.(1997)Selby, Drapkin, Reinberg, and Sancar]{Selby1997}
Selby,~C.~P., Drapkin,~R., Reinberg,~D., and Sancar,~A., 1997.
\newblock {RNA polymerase II stalled at a thymine dimer: footprint and effect
  on excision repair}.
\newblock \emph{Nucleic Acids Res.}, 25\penalty0 (4):\penalty0 787--793.
\newblock \url{http://dx.doi.org/10.1093/nar/25.4.787}.

\bibitem[Shaw et~al.(2003)Shaw, Zia, and Lee]{Shaw2003}
Shaw,~L.~B., Zia,~R. K.~P., and Lee,~K.~H., 2003.
\newblock {Totally asymmetric exclusion process with extended objects: A model
  for protein synthesis}.
\newblock \emph{Phys. Rev. E}, 68\penalty0 (2):\penalty0 021910.
\newblock \url{http://dx.doi.org/10.1103/PhysRevE.68.021910}.

\bibitem[Shaw et~al.(2004{\natexlab{a}})Shaw, Kolomeisky, and Lee]{Shaw2004a}
Shaw,~L.~B., Kolomeisky,~A.~B., and Lee,~K.~H., 2004{\natexlab{a}}.
\newblock {Local inhomogeneity in asymmetric simple exclusion processes with
  extended objects}.
\newblock \emph{J. Phys. A}, 37\penalty0 (6):\penalty0 2105--2113.
\newblock \url{http://dx.doi.org/10.1088/0305-4470/37/6/010}.

\bibitem[Shaw et~al.(2004{\natexlab{b}})Shaw, Sethna, and Lee]{Shaw2004}
Shaw,~L.~B., Sethna,~J.~P., and Lee,~K.~H., 2004{\natexlab{b}}.
\newblock {Mean-field approaches to the totally asymmetric exclusion process
  with quenched disorder and large particles}.
\newblock \emph{Phys. Rev. E - Stat. Nonlinear, Soft Matter Phys.}, 70\penalty0
  (2 1):\penalty0 1--7.
\newblock \url{http://dx.doi.org/10.1103/PhysRevE.70.021901}.

\bibitem[Tkacik et~al.(2008)Tkacik, Gregor, and Bialek]{Tkacik2008}
Tkacik,~G., Gregor,~T., and Bialek,~W., 2008.
\newblock {The role of input noise in transcriptional regulation.}
\newblock \emph{PLoS One}, 3\penalty0 (7):\penalty0 e2774.
\newblock \url{http://dx.doi.org/10.1371/journal.pone.0002774}.

\bibitem[van Kampen(1992)]{vanKampen1992}
van Kampen,~N., 1992.
\newblock \emph{{Stochastic Processes in Physics and Chemistry}}.
\newblock North-Holland Personal Library, Amsterdam.

\bibitem[Xiao et~al.(2016)Xiao, Chen, and Liu]{Xiao2016}
Xiao,~S., Chen,~X., and Liu,~Y., 2016.
\newblock {Totally asymmetric simple exclusion process with a single defect
  site on boundaries}.
\newblock \emph{Int. J. Mod. Phys. B}, 30\penalty0 (14):\penalty0 1650083.
\newblock \url{http://dx.doi.org/10.1142/S0217979216500831}.

\bibitem[Zia et~al.(2011)Zia, Dong, and Schmittmann]{Zia2011}
Zia,~R. K.~P., Dong,~J.~J., and Schmittmann,~B., 2011.
\newblock {Modeling Translation in Protein Synthesis with TASEP: A Tutorial and
  Recent Developments}.
\newblock \emph{J. Stat. Phys.}, 144\penalty0 (2):\penalty0 405--428.
\newblock \url{http://dx.doi.org/10.1007/s10955-011-0183-1}.

\end{thebibliography}
\bibliographystyle{my_plainnat}
\footnotesize
\setlength{\bibsep}{0pt}

\end{document}